\documentclass[a4paper]{article}
\usepackage{amsmath}
\usepackage{amsfonts}
\usepackage{amssymb}
\usepackage{graphics,graphicx}
\usepackage{cite}
\usepackage{latexsym}
\usepackage{textcomp,color}
\usepackage{url, hyperref}

\title{Weyl geometry and cosmological particle creation}
\author{Victor\,A.\,Berezin and Vyacheslav\,I.\,Dokuchaev 
\\ \\ \it Institute for Nuclear Research of the RAS,
\\ prospekt 60-letiya Oktyabrya 7a, Moscow, 117312 Russia}
\begin{document}
\date{\empty} 
\maketitle
\begin{abstract}
These are lectures for students at the summer school conducted by the Faculty of Fundamental Sciences, Bauman Moscow State Technical University at 2022. 
\end{abstract}

\tableofcontents

\section{Differential geometry}

The differential geometry starts with  the definition of the manifold which is just the set of points covered by patches with the coordinate systems $x^\mu$, $ {x'}^{\,\mu'}$ and so on. In the overlapping regions there exist smooth enough and invertible functions  $x^\mu=x^\mu({x'}^{\,\mu'})$.

The main objects are scalars, vectors and tensors, they are defined by the following transformation laws.

Scalars:
\begin{equation}\label{Scalars}
\varphi(x)=\varphi(x').
\end{equation}
Vectors: 
\begin{equation}\label{Vectors}
l^{\mu'}(x')=\frac{\;\;\partial {x'}^{\mu'}}{\partial x^\mu}l^\mu(x), \quad
l_{\mu'}(x')=\frac{\partial x^\mu}{\;\;\partial {x'}^{\mu'}}l_\mu(x).
\end{equation}
Tensors: 
\begin{equation}\label{Tensors}
A^{\mu'\nu'}(x')=\frac{\;\;\partial {x'}^{\mu'}}{\partial x^\mu} \frac{\;\;\partial {x'}^{\nu'}}{\partial x^\nu}A^{\mu\nu}(x),
\end{equation}
\begin{equation}\label{Tensors2}
A_{\mu'\nu'}(x')=\frac{\partial x^\mu}{\;\;\partial {x'}^{\mu'}} \frac{\partial x^\nu}{\;\;\partial {x'}^{\nu'}}A_{\mu\nu}(x),
\end{equation}
\begin{equation}\label{Tensors3}
A^{\mu'}_{\nu'}(x')=\frac{\;\;\partial {x'}^{\mu'}}{\partial x^\mu} \frac{\partial x^\nu}{\;\;\partial {x'}^{\nu'}}A^{\mu}_{\nu}(x)\ldots.
\end{equation}
The ellipsis means that there may be more than two indices. In the above formulas (as well as in many others below) it is assumed the summation over the same indices appeared up and down (the so called Einstein's rule). Main feature of these objects is that  the ``absence'' ($=$ ``the existence'') does not depend on the choice of the coordinate system (their zero values are absolute).

In what follows we will be dealing only with the four-dimensional space-times. Thus,  the indices will take values $(0,1,2,3)$, the index ``$0$'' being reserved for the time-like coordinate. 

Let us consider the simplest case --- the Minkowski space-time of Special Relativity. In 1907 Hermann Minkowski (1864--1909) introduced the notion ``interval'', $ds$, between the neighbouring points. Its square, in the Cartesian (orthogonal) coordinates equals 
\begin{equation}\label{ds2}
ds^2=dt^2-d{\vec x}^2,
\end{equation}
what is a generalization of the Pythagoras' theorem. Such an interval is invariant under the Lorentz transformation of Special Relativity, i.\,e., remains the same for any inertial observer. Transformed to the general (curvilinear) coordinates $t'(t,{\vec x})$, $\vec x\,'={\vec x}\,'(t,{\vec x})$, it becomes
\begin{equation}\label{ds2b}
	ds^2=g_{\mu\nu}(x')dx'^\mu dx'^\nu.
\end{equation}
In the general space-time, $g_{\mu\nu}(x)$ is called the metric tensor, and its role is of most importance. It is assumed that the metric tensor is non-degenerate, i.\,e., 
\begin{equation}\label{detg}
g={\rm det}(g_{\mu\nu})=||g_{\mu\nu}||\neq0,
\end{equation}
and it is symmetric, $g_{\mu\nu}=g_{\nu\mu}$. Therefore, there exists  its reverse, $g^{\mu\nu}$,
\begin{equation}\label{reverse}
g^{\mu\lambda}g_{\lambda\nu}=\delta^\mu_\nu,
\end{equation}
where $\delta^\mu_\nu$ is the Kronecker symbol ($=$ unit tensor).

The metric tensor $g_{\mu\nu}$ and its inverse $g^{\mu\nu}$ allow us to lower and raise indices, so from now on $l_\mu=g_{\mu\nu}l^\nu$ and $l_\mu\equiv g^{\mu\nu}l_\nu$ will denote the same vector, but differently displayed (the formulas for tensors are evident). Remember the Einstein's rule!

\subsection{Differentiation}

The subtitle ``Differential geometry'' implies that we have to introduce somehow the differentiation procedure. The general definition is the following: the differentiation is any linear operator obeying the Leibniz rule. The linearity means that the operator $D$, acting on the linear combination of two functions $\alpha A(x)+\beta B(x)$, $\alpha,\beta=const$, gives us
\begin{equation}\label{calD}
{\cal D}(\alpha A(x)+\beta B(x))=\alpha({\cal D}A(x)) +\beta({\cal D}B(x)),
\end{equation}
while the Leibniz rule tells us how to deal with the product of two functions,
\begin{equation}\label{product}
{\cal D}(A(x)B(x)) = ({\cal D}A(x))B(x)+A(x)({\cal D}B(x)).
\end{equation}

Everybody knows from the course of the mathematical analysis that the ``ordinary differential'' $dA(x)$ is the linear part of the increment of $A(x)$, namely,
\begin{equation}\label{increment}
A(x\!+\!dx)-A(x)\!=\!dA(x)+\ldots=\!A(x)_{,\mu}dx^\mu+\ldots,
\end{equation}
where $A(x)_{,\mu}\equiv \partial A/\partial x^\mu$ is the partial derivative. Evidently, for a scalar field, $\varphi(x+dx)-\varphi(x)$ is a scalar. But for the vector or tensor fields it is not so, because we have to compare their values in different points with different laws of transformations. We need to transfer somehow the vector in one point to the vector (!) in the neighboring point along some special path. Such a procedure is called the parallel transfer.

In the Minkowski space-time	(4-dim) or in the Euclidean space (3-dim) all the points are equivalent, and the parallel transport of any vector  $l^\mu$ in Galilean (Cartesian) coordinates is, by definition, simply 
\begin{equation}\label{lmu}
dl^\mu=0,
\end{equation}
since the metric tensor does not depend on $x$. In the curvilinear coordinates it is not so.

Let us introduce some new differential operator, $\cal D$, and demand that the vector after transferring remains the vector. For the parallel transfer, by definition, 
\begin{equation}\label{lmu2}
{\cal D}l^\mu=0.
\end{equation}
(For the scalars, ${\cal D}\varphi=d\varphi$.)

The linearity condition requires that 
\begin{equation}\label{Dlmu}
	{\cal D}l^\mu= dl^\mu+\Gamma^\mu_{\nu\lambda}l^\lambda dx^\nu.
\end{equation}

\subsection{Connections}
The coefficients $\Gamma^\mu_{\nu\lambda}$ are called connections. The requirement for vectors to remain vectors leads to the following transformation law
\begin{equation}\label{Gamma2a}
	\Gamma^{\mu'}_{\nu'\lambda'} = \frac{\partial^2 x^\mu}{\partial {x'}^{\nu'}\partial {x'}^{\lambda'}}
	\frac{\partial {x'}^{\mu'}}{\partial x^\mu} + \frac{\partial x^\lambda}{\;\;\partial {x'}^{\lambda'}}
	\frac{\partial x^\nu}{\;\;\partial {x'}^{\nu'}}
	\frac{\partial x^{\mu'}}{\partial x^\mu}
	\Gamma^\mu_{\nu\lambda} 
\end{equation}
This new differential is called ``the covariant differential''. The corresponding covariant derivative for the vector field $l^\mu$ equals
\begin{equation}\label{Dlmu2}
\nabla_\lambda l^\mu\stackrel{                                                                                                                                                                                                                                                                                                                                                                                                                                                                                                                                                                                                                                                                                                                                                                                                                                                                                                                                                                                                                                                                                                                                                                                                                                                                                                                                                                                                                                                                                                                                                                                                                    {\mathrm{def}}}{=} l^\mu_{\;,\lambda} +\Gamma_{\lambda\nu}^\mu l^\nu.
\end{equation}
And what about ${\cal D}p_\mu$? The Leibniz rule will help us because ${\cal D}p_\mu l^\mu$ is a scalar. One has 
\begin{equation}\label{DPmu}
{\cal D}(p_\mu l^\mu)=l^\mu({\cal D}p_\mu)+p_\mu({\cal D}l^\mu) =l^\mu({\cal D}p_\mu)+p_\mu(dl^\mu+\Gamma^\mu_{\lambda\nu}l^\nu dx^\lambda),
\end{equation}
But, ${\cal D}(p_\mu l^\mu)=d(p_\mu l^\mu)$, hence
\begin{equation}\label{DPmu2}
{\cal D}(p_\mu l^\mu)=l^\mu(d p_\mu)+p_\mu(d l^\mu),
\end{equation}
and one obtains from this, that 
\begin{equation}\label{DPmu3}
{\cal D}p_\mu=dp_\mu-\Gamma^\nu_{\lambda\mu}p_\nu dx^\lambda,
\end{equation}
\begin{equation}\label{DPmu4}
\nabla_\lambda p_\mu=p_{\mu,\lambda} -\Gamma_{\lambda\mu}^\nu p_\nu.
\end{equation}
The Leibniz rule also helps us in establishing the differentiation of tensors if one remembers that the product of two vectors $l^\mu p^\nu$ is a tensor. We will show below the final result 
\begin{eqnarray}\label{Amunu1}
\nabla_\lambda A^{\mu\nu}&=&A^{\mu\nu}_{\phantom{ab},\lambda} + \Gamma^\mu_{\lambda\sigma}A^{\sigma\nu} + \Gamma^\nu_{\lambda\sigma}A^{\mu\sigma}, \\
\nabla_\lambda A_{\mu\nu}&=&A_{\mu\nu,\lambda} - \Gamma^\sigma_{\lambda\mu}A_{\sigma\nu} - \Gamma^\sigma_{\lambda\nu}A_{\mu\sigma}, \\
\nabla_\lambda A^\mu_\nu&=&A^\mu_{\nu,\lambda} + \Gamma^\mu_{\lambda\sigma}A^\sigma_\nu - \Gamma^\sigma_{\lambda\nu}A^\mu_\sigma.
\end{eqnarray}
What could happen if some vector $l^\mu$ would undergo the parallel transfer along an infinitesimal closed contour around some point? In general, its length ($l^\mu l_\mu$) and orientation will be changed. This gives us the knowledge about the neighborhood  of the point in question (topological property). It appears that the behavior of the vector is defined by the tensor of fourth rank, 
$R^{\mu}_{\phantom{\mu}\nu\lambda\sigma}$, called the curvature tensor,
\begin{equation}\label{curvature}
R^{\mu}_{\phantom{\mu}\nu\lambda\sigma}=\frac{\partial \Gamma^\mu_{\nu\sigma}}{\partial x^\lambda}-\frac{\partial \Gamma^\mu_{\nu\lambda}}{\partial x^\sigma}+\Gamma^\mu_{\varkappa\lambda}\Gamma^\varkappa_{\nu\sigma}-\Gamma^\mu_{\varkappa\sigma}\Gamma^\varkappa_{\nu\lambda}.
\end{equation}
Note, that, by construction, $ R^{\mu}_{\phantom{\mu}\nu\lambda\sigma}=-R^{\mu}_{\phantom{\mu}\nu\sigma\lambda}$. If $R^{\mu}_{\phantom{\mu}\nu\lambda\sigma}=0$, the space-time (or space) is called flat. By summing over $\mu=\lambda$, one obtains the new tensor of the second rank, the Ricci tensor,
\begin{equation}\label{Ricci}
R_{\nu\sigma}=R^{\lambda}_{\nu\lambda\sigma}.
\end{equation}
Such an operation (as well as its result) is called ``the convolution''. The convolution of the Ricci tensor gives us the curvature scalar
\begin{equation}\label{scalar}
R=R^{\lambda}_\lambda=g^{\nu\sigma}R_{\nu\sigma}.
\end{equation}
This completes the description of the differential geometry.

So, in order to construct any differential geometry, one needs to know the metric tensor $g_{\mu\nu}(x)$ and the connections $\Gamma^\mu_{\nu\lambda}(x)$. But, the connections do not behave like a tensor, and, this is rather inconvenient.

It turns out that the connections can be calculated if one knows three tensors (the trinity), the metric tensor $g_{\mu\nu}(x)$, the torsion $S^\lambda_{\mu\nu}$,
\begin{equation}\label{torsion}
S^\lambda_{\phantom{\mu}\mu\nu}=\Gamma^\lambda_{\mu\nu} -\Gamma^\lambda_{\nu\mu},
\end{equation}
and the nonmetricity $Q_{\lambda\mu\nu}$,
\begin{equation}\label{nonmetricity}
Q_{\lambda\mu\nu}= \nabla_\lambda g_{\mu\nu}.
\end{equation}
Here we write down the final result which can be obtained by solving the above linear partial differential equations. Namely, 
\begin{equation}\label{nonmetricity2}
\Gamma^\lambda_{\mu\nu}=C^\lambda_{\phantom{\mu}\mu\nu} + K^\lambda_{\phantom{\mu}\mu\nu} +L^\lambda_{\phantom{\mu}\mu\nu},
\end{equation}
where $C^\lambda_{\phantom{\mu}\mu\nu}$ are famous Christoffel symbols,
\begin{equation}\label{Christoffel}
C^\lambda_{\phantom{\mu}\mu\nu}=\frac{1}{2}g^{\lambda\kappa} (g_{\kappa\mu,\nu}+g_{\kappa\nu,\mu}-g_{\mu\nu,\kappa}),
\end{equation}  
and
\begin{equation}\label{K}
K^\lambda_{\phantom{\mu}\mu\nu}=\frac{1}{2} (S^\lambda_{\phantom{\mu}\mu\nu}-S^{\phantom{\mu}\lambda}_{\mu\phantom{\mu}\nu} -S^{\phantom{\mu}\lambda}_{\nu\phantom{\mu}\mu}),
\end{equation}  
\begin{equation}\label{L}
L^\lambda_{\phantom{\mu}\mu\nu}=\frac{1}{2} (Q^\lambda_{\phantom{\mu}\mu\nu}-Q^{\phantom{\mu}\lambda}_{\mu\phantom{\mu}\nu} -Q^{\phantom{\mu}\lambda}_{\nu\phantom{\mu}\mu}).
\end{equation}

\section{Riemannian geometry}

The famous Riemannian geometry appears the most simple, both torsion and nonmetricity  are zero,
\begin{equation}\label{SQ}
S^\lambda_{\mu\nu}=0, \quad Q^\lambda_{\phantom{\mu}\mu\nu}=0.
\end{equation}  
That is 
\begin{equation}\label{Gamma2}
\Gamma^\lambda_{\mu\nu}=\Gamma^\lambda_{\nu\mu}=C^\lambda_{\mu\nu}
=\frac{1}{2}g^{\lambda\kappa}(g_{\kappa\mu,\nu} +g_{\kappa\nu,\mu} +g_{\mu\nu,\kappa}),
\end{equation}
\begin{equation}\label{nabla2}
\nabla_\lambda g_{\mu\nu}=g_{\nu\mu;\lambda}=0.
\end{equation}
By semicolon, ``;'', we denote the covariant derivative, when the connections are the Christoffel symbols.

In the Riemannian geometry the curvature tensor, $R^{\mu}_{\phantom{\mu}\nu\lambda\sigma}$, acquires  some additional algebraic symmetries,
\begin{equation}\label{symmetries}
R_{\mu\nu\lambda\sigma}=R_{\lambda\sigma\mu\nu} = -R_{\nu\mu\lambda\sigma}=-R_{\mu\nu\sigma\lambda}
\end{equation}
\begin{equation}\label{symmetries2}
R^\mu_{\phantom{\mu}\nu\lambda\sigma} +R^\mu_{\phantom{\mu}\sigma\nu\lambda} +R^\mu_{\phantom{\mu}\lambda\sigma\nu} =0,
\end{equation}
and obeys the Bianchi identities
\begin{equation}\label{symmetries2b}
R^\mu_{\phantom{\mu}\nu\lambda\sigma;\kappa} +R^\mu_{\phantom{\mu}\nu\kappa\lambda;\sigma} + R^\mu_{\phantom{\mu}\nu\sigma\kappa;\lambda}=0.
\end{equation}
Also the Ricci tensor is symmetric
\begin{equation}\label{Riccisymm}
R_{\mu\nu}=R_{\nu\mu}.
\end{equation}

The main and famous application of the Riemannian geometry to theoretical physics is, of course, General Relativity. But before coming to this , we will make some lyrical digression.

The modern theoretical physics is based on the so called Least Action Principle. The invention of the least action principles is associated with the names of Pierre de Fermat (1601--1665), Pierre-Louis Mareau de Moupetuis  (1698--1759), Leonard Euler (1707--1783) and Joseph Louis Lagrange (1736--1813).

Fermat was not a professional mathematician, but maid many great achievements. Among them, he explained the reflection of the light rays using the  least action principle, but his works were not published during his life. Moupetuis rediscovered this and introduced the very notion ``action''. Euler and Lagrange developed the corresponding mathematical formalism known now  as the variational calculus. How does it work? 

First of all, one should choose suitable dynamical variables describing the system under consideration. The next step is the construction of some function of these variables and its derivatives, taking into account ``the first principles'', i.\,e., space-time symmetries, covariance ($=$ independence of the choice of the coordinate system), specific symmetries and other properties of the problem and so on.  Such a function is called ``the Lagrangian'', its choice, of course, is not unique.

With the Lagrangian at hand, one should integrate it over the space-time region inside the some boundary. And this is the  action functional we are looking for. The least action principle reads as follows. ```Among all the trajectories of dynamical variables, the ``true'' ones are that provide the minimum of the action integral''.

Let us denote the set of the dynamical variables by $\psi$, and the action integral by $S$. To find a minimum (extremum) of $S$ we should make a variation of the dynamical variable, $\delta\psi$, and then put the corresponding variation  $\delta S$ to zero. The variations $\delta\psi$ of course, can be arbitrary small (infinitesimal) so the rules of operation are the same as with the derivatives. The only difference is that the dynamical variables $\psi_1$ and $\psi_2$, $\delta\psi=\psi_2(x)-\psi_1(x)$ are taken in the same point. Thus, if $\psi$ is, say, a vector, then $\delta\psi$ is also a vector. Evidently the variation commutes with the partial derivative, i.\,e., $\delta(\psi_\mu)=(\delta\psi)_\mu$. It is most important to note that the variations $\delta\psi$ will inevitably appear at the boundary surface of integration, $\Sigma$. It is required that this variations should vanish. Because of this one may read in the textbooks about ``the least action principle with the fixed ends''. Below are some simple, but useful, examples of how all this works in practice.

\subsection{Classical examples}

\subsubsection{One dynamical variable}

Let us consider a system with only one dynamical variable  $x(t)$ and the Lagrangian ${\cal L}={\cal L}(x(t),\dot x(t))$, $\dot x$ --- time derivative of $x$. Then the action is
\begin{equation}\label{Lagrangian}
 S=\!\int_a^b\!{\cal L}(x(t),\dot x(t),t)\,dt.
\end{equation}
The variation of $S$ equals
\begin{eqnarray}\label{deltaS2b}
\delta S&=&\!\int_a^b\!(\delta{\cal L})\,dt 
= \!\int_a^b\!\left\{\frac{\partial{\cal L}}{\partial x}(\delta x)+ \frac{\partial{\cal L}}{\partial \dot x}(\delta\dot x) \right\}dt 
\nonumber \\
&=& \!\int_a^b\!\left\{\frac{\partial{\cal L}}{\partial x}(\delta x)+ \frac{d}{dt}
\left(\frac{\partial{\cal L}}{\partial \dot x}\delta x\right) -
\left(\frac{d}{dt}\frac{\partial{\cal L}}{\partial \dot x}\right)(\delta x)
\right\}dt. 
\end{eqnarray}
The second term is the full derivative and can be integrated over,
\begin{equation}\label{deltaS3}
\delta S=\!\frac{\partial{\cal L}}{\partial\dot x}(\delta x)\Big|_a^b+ \!\int_a^b\!
\left\{\frac{\partial{\cal L}}{\partial \dot x} -
\left(\frac{d}{dt}\frac{\partial{\cal L}}{\partial \dot x}\right)\right\}(\delta x)dt.
\end{equation}
Since $\delta x(a)=\delta x(b)=0$ (fixed ends!) and $\delta x$ in the integrand is arbitrary, we arrive at the famous Euler-Lagrange equation
\begin{equation}\label{EL}
\frac{\partial{\cal L}}{\partial x}-  
\frac{d}{dt}\frac{\partial{\cal L}}{\partial \dot x}=0.
\end{equation}

\subsubsection{Scalar field}

The second example is the system with continuous ``number'' of dynamical variables, let it be the scalar field  $\varphi(x^\mu)$, $\mu=0,1,2,3$. The action $S$ is
\begin{equation}\label{ActionScalar}
S=\!\int\!{\cal L}\left(\varphi(x),\varphi_{,\mu}(x)\right)\sqrt{-g}\,d^4x,
\end{equation}
here $g={\rm det}(g_{\mu\nu})$ $<0$, $\varphi_{,\mu}={\partial \varphi(x)}/{\partial x^\mu}$, $\Omega$ is the volume of integration with the boundary $\Sigma$. Then
\begin{eqnarray}\label{deltaSscalar}
\delta S&=&\!\!\!\int\limits_\Omega\!\left\{\frac{\partial{\cal L}}{\partial \varphi}(\delta\varphi)+ \frac{\partial{\cal L}}{\partial\varphi_{,\mu}}(\delta \varphi_{,\mu}) \right\}\sqrt{-g}\,d^4x 
\\
&=&\!\!\!\int\limits_\Omega\!\!\left\{\!\frac{\partial{\cal L}}{\partial \varphi}\sqrt{-g}(\delta\varphi)\!+\! \left(\frac{\partial{\cal L}}{\partial\varphi_{,\mu}}\sqrt{-g}(\delta\varphi)\right)_{\!,\mu}\!\!-\!
\left(\!\frac{\partial{\cal L}}{\partial\varphi_{,\mu}}\sqrt{-g}\right)_{\!,\mu}\!\! (\delta\varphi)\! \right\}\!d^4x. \nonumber 
\end{eqnarray}
Now, we will make use of remarkable Stokes' theorem: for any vector $l^\mu$,
\begin{equation}\label{ActionScalar2}
\!\!\int\limits_\Omega l^\mu_{\:\:,\mu}=\!\int\limits_\Sigma l^\mu dS_\mu.
\end{equation}
We do not intend to explain what $dS_\mu$ is, it is essential for us now that the volume integral of the full derivative is converted into the surface integral. Thus, we have
\begin{equation}\label{deltaActionScalar}
\delta S= \!\!\int\limits_\Sigma\!\!\frac{\partial{\cal L}}{\partial\varphi_{,\mu}}\sqrt{-g}(\delta\varphi)dS_\mu+\!\!\int\limits_\Omega\!\!\left\{\!\frac{\partial{\cal L}}{\partial \varphi}\!-\!\frac{1}{\sqrt{-g}} \left(\!\!\frac{\partial{\cal L}}{\partial\varphi_{,\mu}}\sqrt{-g}\right)_{\!\!,\mu}\! \right\}  \!(\delta\varphi)\sqrt{-g}\,d^4x.
\end{equation}
The surface integral vanishes because $\delta\varphi=0$ on $\Sigma$ (fixed ends!). We are left with the following Euler-Lagrange equation,
\begin{equation}\label{deltaELscalar}
\frac{\partial{\cal L}}{\partial\varphi}-  
\left(\frac{\partial{\cal L}}{\partial\varphi_{,\mu}}\right)_{\!;\mu}=0.
\end{equation}

\section{General Relativity}

Let us come, at last, to General Relativity. This incredible theory combines physics and geometry and has two parents, Albert Einstein (physics) and David Hilbert (mathematics). 

Einstein's idea: matter fields make the space-time curved and this very curvature is that we feel as the gravitation. The Hilbert's idea: all the physical equations may be deduced from the least action principles. By the end of 1915 both programs were, actually, fulfilled.

According to Einstein, the left-hand-side should be pure geometrical, while the right-hand-side should be proportional to the energy-momentum tensor, $T_{\mu\nu}$, of the matter fields as a source. According to Hilbert, the total action, $S_{\rm tot}$, should be the sum of the gravitational (geometrical) part, $S_{\rm grav}$, and the matter part $S_{\rm m}$, i.\,e., 
\begin{equation}\label{Stot}
S_{\rm tot}=S_{\rm grav}+S_{\rm m}.
\end{equation}
Both of them kept in mind the Riemannian geometry, so, the only dynamical variable could be the metric tensor $g_{\mu\nu}$. By definition,
\begin{equation}\label{Stot2}
\delta S_{\rm m}\stackrel{\mathrm{def}}{=}
-\frac{1}{2}\int\! T_{\mu\nu}(\delta g^{\mu\nu})\sqrt{-g}\,d^4x
=\frac{1}{2}\int\! T^{\mu\nu}(\delta g_{\mu\nu})\sqrt{-g}\,d^4x.
\end{equation}
For the gravitational Lagrangian, Hilbert made the simplest possible choice --- curvature scalar $R$. Thus, the so  called Hilbert action equals
\begin{equation}\label{Hilbertaction}
S_{\rm grav}=S_{\rm H}=-\frac{1}{16\pi G} \int\! R\sqrt{-g}\,d^4x,
\end{equation}
where $G$ is the  Newton's gravitational constant (remember that $c=1$). This coefficient is chosen in order to have the correct non-relativistic limit and the minimum (not maximum) of the action (the ``minus sign'').

Thus
\begin{equation}\label{deltaH}
\delta\!\int\!R\sqrt{-g}\,d^4x=8\pi G\! \int\! T_{\mu\nu}(\delta g^{\mu\nu})\sqrt{-g}\,d^4x.
\end{equation}
Since $R=g^{\mu\nu}R_{\mu\nu}$  and $\delta\sqrt{-g}=-(1/2)\sqrt{-g}g^{\mu\nu}(\delta g^{\mu\nu}$ (famous formulas from the textbooks), one readily has
\begin{equation}\label{deltaH2}
\delta\!\int\!R\sqrt{-g}\,d^4x= \!\int\!\Bigl\{ g^{\mu\nu}(\delta R_{\mu\nu})+(R_{\mu\nu}- \frac{1}{2}g_{\mu\nu}R)(\delta g^{\mu\nu})\!\Bigr\} \sqrt{-g}\,d^4x.
\end{equation}
For $\delta R_{\mu\nu}$ there exists the remarkable formula by Palatini\cite{Palatini}, found in 1919,   
\begin{equation}\label{deltaH2b}
\delta R_{\mu\nu}=(\delta\Gamma^\lambda_{\mu\nu})_{;\lambda} -(\delta\Gamma^\lambda_{\mu\lambda})_{;\nu}
\end{equation}
(more precisely, the metric covariant derivative ``;'' should be replaced by the general one, $\nabla$). Everybody can easily check this, starting from the definition of Ricci tensor and remembering that $\delta\Gamma^\lambda_{\mu\nu}$ is a tensor (unlike $\Gamma^\lambda_{\mu\nu}$ itself). Then, noticing that 
\begin{equation}\label{g5}
g^{\mu\nu}(\delta\Gamma^\lambda_{\mu\nu})_{;\lambda} \sqrt{-g} = \left(g^{\mu\nu}(\delta\Gamma^\lambda_{\mu\nu})\right)_{;\lambda} \sqrt{-g} = \left(g^{\mu\nu}(\delta\Gamma^\lambda_{\mu\nu})\sqrt{-g}\right)_{,\lambda} 
\end{equation}
is a full derivative and using the Stokes' theorem, one gets immediately the famous Einstein equation
\begin{equation}\label{EinsteinEq}
R_{\mu\nu} - \frac{1}{2}g_{\mu\nu}R=8\pi G T_{\mu\nu}.
\end{equation}
If David Hilbert would know the Palatini formula in 1915! 

\section{Weyl Geometry}

The triumph of the geometrical description of the gravitational interactions forced scientists to search for the unification of all known at the time interactions --- gravity$+$electromagnetism --- under the auspices of geometry. Hermann Weyl offered such a geometry.

\subsection{Electromagnetic field and gauge invariance}

In order to understand, how it was done we have to remind some facts concerning the classical electrodynamics. Electromagnetic field are described by the vector-potential $A_\mu$. The electric and magnetic strengths are the components of the anti-symmetric second rank tensor $F_{\mu\nu}$,
\begin{equation}\label{F}
F_{\mu\nu}=A_{\nu,\mu} - A_{\mu,\nu} =A_{\nu;\mu} - A_{\mu;\nu}.
\end{equation}
The first pair of Maxwell equations is just the identity
\begin{equation}\label{M1}
F_{\mu\nu;\sigma}+F_{\sigma\mu;\nu} +F_{\nu\sigma;\mu}=0,
\end{equation}
while the second pair has the form
\begin{equation}\label{M2}
F^{\mu\nu}_{;\nu}= \frac{(F^{\mu\nu}\sqrt{-g})_{,\nu}}{ \sqrt{-g}}=-4\pi j^\mu,
\end{equation}
where $j^\mu$ is the electric current density. It easy to verify that the electric charge is conserved, i.\,e., $j^\mu_{\;;\mu}=0$.

We see that the electromagnetic  vector-potential $A_\mu$ does not enter the Maxwell equations, only $F_{\mu\nu}$ are present (and measurable). The  tensor $F_{\mu\nu}$ is, obviously, invariant under the transformation $A_\mu \; \rightarrow \; A_\mu+\alpha_{,\mu}$, where $\alpha$ is an arbitrary scalar field. This the so called gauge transformation and, correspondingly, gauge invariance. It is known from the course of the electrodynamics, that the action for the system ``single charged particle of  mass $m$ $+$ electromagnetic field'' has the form
\begin{equation}\label{M2b}
S=-m\!\int\!ds-\!\int\!A_\mu j^\mu\sqrt{-g}\,d^4x -\frac{1}{16\pi}\!\int\!F_{\mu\nu}F^{\mu\nu}\sqrt{-g}\,d^4x.
\end{equation}
It is clear that the second term is not manifestly gauge invariant. And it does not need to be! But $\delta S(=0)$ --- does! Let us consider an infinitesimal gauge transformation $\delta\alpha$. The corresponding response,  $\delta S$, is
\begin{equation}\label{deltaS2c}
	\delta S=-\!\int\!(\delta A_\mu) j^\mu\sqrt{-g}\,d^4x = -\!\int\!j^\mu(\delta\alpha)_{,_\mu} \sqrt{-g}\,d^4x.
\end{equation}
Separating the full derivative and neglecting it, one gets
\begin{equation}\label{deltaS3b}
\delta S=\!\int\!\frac{(j^\mu\sqrt{-g})_{,\mu}}{\sqrt{-g}}(\delta \alpha)\sqrt{-g}\,d^4x = \!\int\!j^\mu_{\;;\mu}(\delta\alpha)\sqrt{-g}\,d^4x=0.
\end{equation}
We obtained the electric charge conservation law, $j^\mu_{\;;\mu}=0$. This is the self-consistency condition.

\subsection{Weyl geometry versus Riemannian geometry}

Hermann Weyl\cite{Weyl} put forward the following pure physical idea: in the course of the parallel transfer of some rod ($=$ vector), its length is changing (unlike in the Riemannian geometry). It seems quite natural in the unified theory because the rods are made of the charged particles. Thus,
\begin{equation}\label{nabla3}
\nabla_\lambda g_{\mu\nu}\neq0.
\end{equation}
Weyl's choice was the simplest one:
\begin{equation}\label{nabla3b}
Q_{\lambda\mu\nu}=\nabla_\lambda g_{\mu\nu}=A_\lambda g_{\mu\nu},
\end{equation}
where $A_\lambda$ is the electromagnetic vector potential. Assuming  $S^\lambda_{\mu\nu}=0$ ($\Gamma^\lambda_{\mu\nu}=\Gamma^\lambda_{\nu\mu}$), one gets
\begin{equation}\label{GammaW}
\Gamma^\lambda_{\mu\nu}=C^\lambda_{\mu\nu}+W^\lambda_{\mu\nu},
\end{equation}
\begin{equation}\label{CW}
C^\lambda_{\mu\nu}=\frac{1}{2}g^{\lambda\kappa} (g_{\kappa\mu,\nu}+g_{\kappa\nu,\mu}-g_{\mu\nu,\kappa}),
\end{equation}
\begin{equation}\label{WW}
W^\lambda_{\mu\nu}=-\frac{1}{2}(A_\mu \delta^\lambda_\nu+ A_\nu \delta^\lambda_\mu -A^\lambda g_{\mu\nu}).
\end{equation}
This is not the end of the story.

\subsection{Conformal transformation}

The change in the length can be compensated by the suitable local conformal transformation, which is the following,
\begin{equation}\label{hats}
ds^2=\Omega^2(x)d\hat s^2,
\end{equation}
\begin{equation}\label{hats2}
g_{\mu\nu}=\Omega^2(x)\hat g_{\mu\nu}, \quad g^{\mu\nu}=\frac{1}{\Omega^2(x)}\hat g^{\mu\nu}.
\end{equation}
Important note: the local conformal transformation does not touch upon the coordinate system, it is simply the change of the measurement units.

The Christoffell symbols $C^\lambda_{\mu\nu}$ are transformed as 
\begin{equation}\label{Christhats}
C^\lambda_{\mu\nu}=\hat C^\lambda_{\mu\nu} +\left(\frac{\Omega_{,\mu}}{\Omega}\delta^\lambda_\nu +\frac{\Omega_{,\nu}}{\Omega}\delta^\lambda_\mu -\hat g^{\lambda\sigma}\frac{\Omega_{,\sigma}}{\Omega}\hat g_{\mu\nu}\right).
\end{equation}
Surprisingly enough, if one would demand
\begin{equation}\label{amuW}
A_\mu=\hat A_\mu+2\frac{\Omega_{,\mu}}{\Omega},
\end{equation}
then
\begin{equation}\label{amuW2}
\Gamma^\lambda_{\mu\nu}=\hat\Gamma^\lambda_{\nu\mu},
\end{equation}
i.\,e.  $A_\mu$ becomes a gauge field when the connections are required  to be  conformal invariant\cite{Penrose,Hooft} --- great discovery! 

Evidently, one has
\begin{equation}\label{Rhat2}
R^\mu_{\phantom{1}\nu\lambda\sigma}=\hat R^\mu_{\phantom{1}\nu\lambda\sigma},
\end{equation}
\begin{equation}\label{Rhat3}
R_{\mu\nu}=\hat R_{\mu\nu},
\end{equation}
\begin{equation}\label{Rhat4}
F_{\mu\nu}=\hat F_{\mu\nu}.
\end{equation}
This is ``the Weyl Geometry''.

\section{Weyl gravity}

It can be easily shown that the Maxwell equations are invariant under the local conformal transformations. Hermann Weyl claimed that in the unified theory the gravitational equations must be conformal invariant too. In analogy with the electromagnetism he decided to include into the Lagrangian only quadratic terms, namely,
\begin{equation}\label{SW4}
S_{\rm W} =\int\!{\cal L_{\rm W}}\sqrt{-g}\,d^4x,
\end{equation}
\begin{equation}\label{Rhat4b}
{\cal L_{\rm W}}=\alpha_1  R_{\mu\nu\lambda\sigma}R^{\mu\nu\lambda\sigma}
+\alpha_2R_{\mu\nu}R^{\mu\nu}+\alpha_3R^2 +\alpha_4F_{\mu\nu}F^{\mu\nu}.
\end{equation}
It is quite clear that the action ${\cal L_{\rm W}}$ is conformal invariant (remember that $\sqrt{-g}=\Omega^4\sqrt{-\hat g}$).

Albert Einstein found some discrepancies between the predictions of Weyl's unified theory and the stability of the atomic spectra (the problem of the ``second time'' and all that). The theory was rejected and almost forgotten.

We will not identify the Weyl vector $A_\mu$ with the electromagnetic vector potential and consider it just as a part of the beautiful conformal invariant geometry. Nowadays, even the fact that the Weyl's Lagrangian $S_{\rm W}$ is quadratic, looks plausible, because the terms quadratic in curvatures (in Riemannian geometry), describe the conformal anomaly in one loop approximation of the quantum field theory\cite{Parker69,GribMam70,ZeldPit71,HuFullPar73,FullParHu74,FullPar74,ZeldStarob,ParkerFulling,GribMamMostep}, the latter being responsible for the vacuum polarization and particle  creation\cite{ZeldStarob2}.

\subsection{Total action}

The total action is
\begin{equation}\label{tot}
S_{\rm tot}=S_{\rm W}+S_{\rm m}.
\end{equation}
Though the Weyl action $S_{\rm W}$, is conformal invariant, the action for the matter fields, $S_{\rm m}$, does not need to be such. But its variation $\delta S_{\rm m}$, is obliged to be conformal invariant. This means that there exists some condition imposed on the form of the matter Lagrangian. Let us find it. In general, the matter Lagrangian  depends on some dynamical variables, $\psi$, describing the matter fields, and on the geometrical variables, $g_{\mu\nu}$ and $A_\mu$.

By definition, 
\begin{eqnarray}\label{Sm3}
\delta S_{\rm m}&=& \!\!-\frac{1}{2}\!\int\!T^{\mu\nu}(\delta g_{\mu\nu})\sqrt{-g}\,d^4x
-\!\int\!\! G^\mu(\delta A_\mu)\sqrt{-g}\,d^4x \nonumber\\
&&\!\!+\int\!\frac{\partial \cal L_{\rm  W}}{\partial\psi} (\delta \psi)\sqrt{-g}\,d^4x,
\end{eqnarray}
where $T^{\mu\nu}$ is the energy-momentum tensor, and $G^\mu$ can be called ``the Weyl current''. By virtue of the Euler-Lagrange equation, the last term equals zero.

Then, consider an infinitesimal conformal transformation
\begin{equation}\label{deltag4}
\delta g_{\mu\nu}=2\Omega\hat g_{\mu\nu}(\delta\Omega)=2g_{\mu\nu}\frac{\delta\Omega}{\Omega},
\end{equation}
\begin{equation}\label{deltaA4}
\delta A_\mu=2\delta\left(\frac{\Omega_{,\mu}}{\Omega}\right)\! = 2\delta(\log\Omega)_{,\mu} =2(\delta(\log\Omega))_{,\mu} =2\left(\frac{\delta\Omega}{\Omega}\right)_{\!,\mu}\!\!.
\end{equation}
Then,
\begin{equation}\label{deltaA4b}
\delta S_{\rm m} =-\frac{1}{2}\int\!T^{\mu\nu}\left(\frac{\delta\Omega}{\Omega} \right)\sqrt{-g}\,d^4x
-\int\! G^\mu\left(\frac{\delta\Omega}{\Omega} \right)\sqrt{-g}\,d^4x =0.
\end{equation}
Separating the full derivative in the second term  and neglecting it (fixed ends!), one gets
\begin{equation}\label{TraceT}
2G^\mu_{\;;\mu}={\rm Trace}(T^{\mu\nu}) .
\end{equation}
This is the self-consistency condition. (Note the metric covariant derivative in the left-hand-side).

\subsection{Perfect fluid}

\begin{eqnarray}\label{perfect}
S_{\rm m}&=& -\!\int\!\varepsilon(X,n)\sqrt{-g}\,d^4x + \int\!\lambda_0(u_\mu u^\mu-1)\sqrt{-g}\,d^4x \nonumber\\
&&+\int\!\lambda_1(n u^\mu)_{;\mu}\sqrt{-g}\,d^4x + \int\!\lambda_2 X_{,\mu}u^\mu\sqrt{-g}\,d^4x,
\end{eqnarray}
with the following dynamical variables: $n(x)$ as the particle number density, $u^\mu(x)$ as the four-velocity vector and $X(x)$ as the auxiliary variable, enumerating the trajectories\cite{Ray,Berezin}. The energy density $\varepsilon(X,n)$ depends on two dynamical variables, and $\lambda_0(x)$, $\lambda_1(x)$ and $\lambda_2(x)$ are the Lagrange multipliers, providing the constraints.

The equations of motion are
\begin{eqnarray}\label{perfect2b}
\delta n:&& -\frac{\partial\varepsilon}{\partial n}-\lambda_{1,\mu}u^\mu=0,\\
\delta u^\mu:&& 2\lambda_0 u^\mu - n\lambda_{1,\mu}+\lambda_2 X_{,\mu}=0, \\
\delta X:&& -\frac{\partial\varepsilon}{\partial X} -(\lambda_2 u^\mu)_{;\mu}=0.
\end{eqnarray}
The constraints are
\begin{equation}\label{eq4}
\delta \lambda_1: \quad (n u^\mu)_{;\mu}=0,
\end{equation}
\begin{equation}\label{eq5}
\delta \lambda_1: \quad (n u^\mu)_{;\mu}=0,
\end{equation}
\begin{equation}\label{eq6}
\delta \lambda_2: \quad X_{,\mu}u^\mu=0.
\end{equation}
The first of them is just the familiar renormalization of the four-velocities, the second one demonstrates the conservation law of the number of particles, while the third constraint tells us that the auxiliary variable $X$ is constant along the trajectory.

The energy-momentum tensor is obtained by varying the metric tensor $\delta g_{\mu\nu}$: $T^{\mu\nu}=\varepsilon g^{\mu\nu}-2\lambda_0 u^\mu u^\nu +n\lambda_{1,\sigma}u^\sigma g^{\mu\nu}
$. By contracting the second of the equations of motion with $u_\mu$, using the constraints and the comparing with the first one, allows us to extract the relation 
\begin{equation}\label{eq6b}
2\lambda_0=-n\frac{\partial\varepsilon}{\partial n}.
\end{equation}
Introducing, then, the hydrodynamical pressure $p$
\begin{equation}\label{p}
p=n\frac{\partial\varepsilon}{\partial n} -\varepsilon,
\end{equation}
we get the familiar energy-momentum tensor for the perfect fluid
\begin{equation}\label{T5b}
T^{\mu\nu}=(\varepsilon+p)u^\mu u^\nu -p g^{\mu\nu}.
\end{equation}
The Weyl geometry may and will cause the modifications of the simple picture presented above. To understand what kind of changes are possible, let us consider the single particle moving in a given space-time.

Everybody knows that in the Riemannian geometry, the only invariant that can be constructed  in order to describe the particle motion is the interval $s$ along its trajectory. Hence, 
\begin{equation}\label{T6}
S_{\rm part} = -m\!\int\!ds=-m\!\int\!\sqrt{g^{\mu\nu}(x)\frac{d x^\mu}{d\tau}\frac{d x^\nu}{d\tau}}d\tau,
\end{equation}
where
$m$ is the particle mass ($u^\mu u_\mu=1$),
$d x^\mu/ds=d x^\mu/d\tau=u^\mu$ --- its four-velocity, and $\tau$ --- the proper time.

The least action principle, $\delta S_{\rm m}=0$, with the fixed ends gives us the geodesic equations (``shortest'' interval),
\begin{equation}\label{u5}
u_{\lambda;\nu}u^\nu=0.
\end{equation}
In the Weyl geometry, however, there exists yet another invariant, 
\begin{equation}\label{B}
B=A_\mu u^\mu,
\end{equation}
and the single particle action may have the more general form,
\begin{equation}\label{Spart}
S_{\rm part} =\!\int\!\!f_1(B)ds \! +\!\int\!\!f_2(B)d\tau = \!\int\!\!\left\{\right.f_1(B)\sqrt{g_{\mu\nu}u^\mu u^\nu} \! + f_2(B)\left.\right\}d\tau.
\end{equation}
The equations of motion become
\begin{equation}\label{f12}
f_1u_{\lambda;\mu}u^\mu=\left( (f_1^{''}+f_1^{''})A_\lambda- f_1^{'}u_\lambda)\right)B_{,\mu}u^\mu + (f_1^{'}+f_2^{'})F_{\lambda\mu}u^\mu,
\end{equation}
where $F_{\lambda\mu}=A_{\mu,\lambda}-A_{\lambda,\mu}$. Since $F_{\lambda\mu}u^\lambda u^\mu\equiv0$ and $u_{\lambda;\sigma}u^\lambda\equiv0$ (due to the normalization), then the contraction with $u^\lambda$ gives us
\begin{equation}\label{f12b}
\left((f_1^{''}+f_2^{''})B-f_1^{'}\right)B_{,\mu}u^\mu=0.
\end{equation}
This is a consistency condition.

How to insert  the interaction with the Weyl vector $A_\mu$ into the perfect fluid Lagrangian? Evidently, the new invariant $B=A_\mu u^\mu$ is tightly linked to the particle number density $n$. Hence, it seems natural to make the replacement $n \;\; \rightarrow \;\; \varphi(B)n$. And how about the particle number conservation law, $(nu^\mu)_{;\mu}=0$?

We have already mentioned that the quadratic in curvatures terms describe the vacuum polarization and may cause the  particle creation from the vacuum fluctuations. Therefore,
\begin{equation}\label{Psi2}
(n u^\mu)_{;\mu}=\Psi({\rm inv}),
\end{equation}
where $\Psi$ is some invariants constructed from the geometrical structures and classical fields which  also are possible sources of the particle production. 

Clearly, particles, can be just counted, point by point, its number should not depend on the metric itself and on the conformal factor, in  particular. 
Let us check this. The local conformal transformation $g_{\mu\nu}= \Omega^2\hat g_{\mu\nu}$ causes the following changes, 
\begin{equation}\label{n2}
n=\frac{\hat n}{\Omega^3},  \;\; u^\mu=\frac{\hat u^\mu}{\Omega},  \;\; \sqrt{-g}=\Omega^4 \sqrt{-\hat g}.
\end{equation}
Thus,
\begin{equation}\label{n2b}
(n u^\mu)_{;\mu}\sqrt{-g}=(n u^\mu)_{,\mu}= (\hat n\hat u^\mu\sqrt{-\hat g})_{,\mu},
\end{equation}
i.\,e., $(n u^\mu)_{;\mu}\sqrt{-g}$ is conformal invariant. The result (though rather simple) is of very importance, since it does not depend at all on the type of geometry or the gravitational Lagrangian.

In the absence of the classical matter fields we have no other choice but the following
\begin{equation}\label{n2c}
\Phi({\rm inv})=\alpha_1'R_{\mu\nu\lambda\sigma}^{\mu\nu\lambda\sigma}
+\alpha_2'R_{\mu\nu}^{\mu\nu}
+\alpha_3'R^2+ \alpha_4'F_{\mu\nu}F^{\mu\nu},
\end{equation}
i.\,e., the Weyl gravitational Lagrangian with ``primed'' coefficients.

The matter action integral $S_{\rm m}$ becomes now,
\begin{eqnarray}\label{perfect2c}
S_{\rm m}=&-& \!\!\int\!\varepsilon(X,\varphi(B)n)\sqrt{-g}\,d^4x + \int\!\lambda_0(u_\mu u^\mu-1)\sqrt{-g}\,d^4x \nonumber \\
&+&\!\!\int\!\lambda_1\left((nu^\mu)_{;\mu}-\Phi({\rm inv})\right)\sqrt{-g}\,d^4x +\!\!\int\!\lambda_2 X_{,\mu}u^\mu\sqrt{-g}\,d^4x.
\end{eqnarray}
Surely, the changes made by us, will influence both Weyl current $G^\mu$, and the energy-momentum tensor, $T^{\mu\nu}$. There are contributions from the energy density $\varepsilon$, we call the $G^\mu[\rm part]$ and $G^{\mu\nu}[\rm part]$, correspondingly, and from the creation function $\Phi({\rm inv})$, we call them $G^\mu[\rm cr]$ and $T^{\mu\nu}[\rm cr]$. The former ones can be easily calculated,
\begin{eqnarray}\label{perfect3}
G^\mu[\rm part]&=&\frac{\varphi'}{\varphi}(\varepsilon+p)u^\mu, \\
T^{\mu\nu}[\rm part]&=&(\varepsilon+p) \left(1-B\frac{\varphi'}{\varphi}\right)u^\mu u^\nu
-pg^{\mu\nu}.
\end{eqnarray}
The creation parts are two lengthy. We will show them in the next Section \ref{Cosmology}, for the cosmological space-times.

\section{Cosmology}\label{Cosmology}

By cosmology we will understand the homogeneous and isotropic space-time described by the Roberson-Walker metric. What is it?

Let us consider the Euclidean 3-dimensional space. Its line element $dl$ is determined by the familiar Pythagoras' theorem
\begin{equation}\label{Pythagor}
	dl^2 = dx^2+dy^2+dz^2 = dr^2+r^2(d\theta^2+\sin^2\theta d\varphi^2).
\end{equation}
The last equality represents the same, but written in the spherical coordinates, $r^2 =x^2+y^2+z^2$ --- radius of the sphere centering at any point. Evidently, this is homogeneous and isotropic. Surely, the most general form looks as follows,
\begin{equation}\label{dl}
	dl^2 =f(r)dr^2+r^2(d\theta^2+\sin^2\theta d\varphi^2).
\end{equation}
Thus we have one unknown function, $f(r)$. So we need only one equation in order to determine it. If one assumes that our 3-geometry is Riemannian (you already know what it means), the 3-dimensional curvature scalar, $K$, should be constant (due to the homogeneity),
\begin{equation}\label{K0}
K=K_0.
\end{equation}
Using general formulas, given above in the first part of the lectures, you may calculate $K$ (as an exercise) and obtain the following equation
\begin{equation}\label{f'}
\frac{2f'}{rf^2}+\frac{2}{r^2}\left(1-\frac{1}{f}\right)=K_0=const,
\end{equation}
which is readily integrated,
\begin{equation}\label{f}
\frac{1}{f}=1-\frac{K_0}{6}r^2+\frac{K_1}{r}.
\end{equation}
The central point, $r=0$, is singular. But it cannot be so --- otherwise, all the points must be singular as well (homogeneity!). Therefore, $K_1=0$, and after the rescaling of $r$ one gets
\begin{equation}\label{dl22}
dl^2=a_0^2\left(\frac{dr^2}{1-kr^2}+r^2(d\theta^2+\sin^2\theta d\varphi^2) \right), \quad k=0,\pm1.
\end{equation}
Because of the homogeneity (again!) it is possible to introduce the global time $t$ and write down the 4-dimensional metric in the form
\begin{equation}\label{dl23}
ds^2=dt^2-a^2(t)\left(\frac{dr^2}{1-kr^2}+r^2(d\theta^2+\sin^2\theta d\varphi^2) \right), \quad k=0,\pm1.
\end{equation}
The time $t$ is called ``the cosmological time''. This is just the Robertson-Walker metric. The only unknown function in the metric tensor is the so called scale factor $a(t)$. 

The Weyl geometry is characterized also by the Weyl vector $A_\mu$. How about it? Due to the very high cosmological symmetry, the Weyl vector may have  only one nonzero component, $A_0(t)$. By the suitable conformal transformation $\Omega(t)$ it can be put zero, since $A(t)= \hat A(t)+2\dot\Omega/\Omega$. This fixes the gauge freedom. Evidently, in cosmology
\begin{equation}\label{dl23b}
F_{\mu\nu}=A_{\nu,\mu}-A_{\mu,\nu}\equiv0.
\end{equation}
Warning! We are not allowed to put $A_\mu=0$ prior to the variation of the action, because $\delta A_\mu\neq0$. So, we should calculate the variation both of the Weyl action, $S_{\rm W}$, and the matter action, $S_{\rm m}$. The result will be the vector gravitational equation with $G^\mu$ in the right-hand-side in the first case, and the definition of $G^\mu$ (in the left-hand-side) in the second case. As already mentioned, the latter consists of two parts, $G^\mu{[\rm part]}$ and $G^\mu{[\rm cr]}$. The first one, that comes from the direct interaction of the particles with the Weyl vector, we presented above. The calculation of the second, that comes from  the particle creation law, is rather cumbersome. Below is the result 
\begin{equation}\label{Gi}
G^i{[\rm cr]}=0,
\end{equation}
\begin{equation}\label{G0cr}
G^0[\rm cr]= -2(2\alpha_1'+\alpha_2')\dot\lambda_1R^0_0
-(\alpha_2'+6\alpha_3')\dot\lambda_1R
-2(\alpha_1'+\alpha_2'+3\alpha_3')\lambda_1\dot R.
\end{equation}
In order to obtain the vector gravitational equation it is sufficient to put $\lambda_1\equiv1$ and to remove primes
\begin{equation}\label{G0cr1}
-2(\alpha_1+\alpha_2+3\alpha_3)\lambda_1\dot R=G^0.
\end{equation}

It is right now that we have the right to put $A_\mu=0$ straight in the Lagrangians. Actually, we will be dealing with the Riemannian geometry, where the Weyl Lagrangian can be rewritten in the following way
\begin{eqnarray}\label{LW3}
{\cal{L}_{\rm  W}}&=& \alpha_1R_{\mu\nu\lambda\sigma}R^{\mu\nu\lambda\sigma} +\alpha_2R_{\mu\nu}R^{\mu\nu} +\alpha_3R^2 +\alpha_4F_{\mu\nu}F^{\mu\nu} \nonumber \\
&=& \alpha C^2+\beta GB+\gamma R^2.
\end{eqnarray}
Here $C^2=C_{\mu\nu\lambda\sigma}C^{\mu\nu\lambda\sigma}$, where $C_{\mu\nu\lambda\sigma}$ is the so called Weyl tensor, defined as the completely traceless part of  the curvature tensor $R_{\mu\nu\lambda\sigma}$,
\begin{eqnarray}\label{LW3b}
C_{\mu\nu\lambda\sigma}&=&R_{\mu\nu\lambda\sigma}
+\frac{1}{2}\left(-R_{\mu\lambda}g_{\nu\sigma} +R_{\mu\sigma}g_{\nu\lambda}
+R_{\nu\lambda}g_{\mu\sigma} -R_{\nu\sigma}g_{\lambda\mu}\right) \nonumber \\
&&+ \frac{1}{6}R(g_{\mu\lambda}g_{\nu\sigma} -g_{\mu\sigma}g_{\nu\lambda}),
\end{eqnarray}
\begin{equation}\label{C2}
C^2=R_{\mu\nu\lambda\sigma}R^{\mu\nu\lambda\sigma} -2R_{\mu\nu}R^{\mu\nu} +\frac{1}{3}R^2.
\end{equation}
Then follows the so called Gauss-Bonnet term, GB,
\begin{equation}\label{C2b}
GB=R_{\mu\nu\lambda\sigma}R^{\mu\nu\lambda\sigma} -4R_{\mu\nu}R^{\mu\nu} +R^2,
\end{equation}
which is the full derivative in the 4-dimensional space-time and does not affect the field equations.

It is easy to calculate the coefficients $\alpha$, $\beta$ and $\gamma$:
\begin{equation}\label{alpha3}
\left\{
\begin{array}{l}	
	\alpha+\beta=\alpha_1  \\
	-2\alpha-4\beta=\alpha_2 \\
	\frac{1}{3}\alpha+\beta+\gamma=\alpha_3. 
\end{array}	
\right.
\end{equation}
The same, surely, true for the ``primed'' version. With this new notations, one has
\begin{equation}\label{alpha3b}
G^i{[\rm cr]}=4\beta\dot\lambda_1R^0_0 -2(\beta+3\gamma)\dot\lambda_1R -6\gamma\lambda_1\dot R.
\end{equation}
The coefficient $\alpha$ does not enter at all, as it should be, since the Weyl tensor is identically zero for any homogeneous and isotropic space-time.

The variation of the metric tensor, $\delta g_{\mu\nu}$ provides us with the left-hand-sides of the tensor gravitational equations (coming from $\delta S_{\rm W}$) and their right-hand-sides, energy momentum tensor
$T^{\mu\nu}$ (coming from $\delta S_{\rm m}$). Note, that in cosmology we need to know only $T^{00}=T^0_0$ and 
$T={\rm Trace}T^{\mu\nu}$, since $T^{0i}=0$ and 
$T^{ij}=T^1_1g^{ij}$, $T^1_1=(1/3)(T-T^0_0)$. Thus,
\begin{equation}\label{T}
T=T[\rm part] +T[\rm cr],
\end{equation}
\begin{equation}\label{T2}
T{[\rm part]}=\varepsilon-3p,
\end{equation}
\begin{eqnarray}\label{Tcr3b}
T{[\rm cr]}&=&\ddot\lambda_1(8\beta'R^0_0 -4\beta'R -12\gamma'R) \nonumber \\
&&-4\dot\lambda_1\left(\beta'\frac{\dot a}{a}(R+2R^0_0)+6\gamma'\dot R +9\gamma'\frac{\dot a}{a}R\right) \nonumber \\
&&-12\lambda_1\gamma'(\ddot R+3\frac{\dot a}{a}\dot R),
\end{eqnarray}
\begin{equation}\label{T002}
T^0_0=T^0_0{[\rm part]}+T^0_0{[\rm cr]},
\end{equation}
\begin{equation}\label{T002b}
	T^0_0{[\rm part]}=\varepsilon,  
\end{equation}
\begin{eqnarray}\label{Tcr3c}
T^0_0{[\rm cr]}&=&8\gamma'\dot\lambda_1\frac{\dot a}{a}R_0^0 -4(\beta'+3\gamma')\dot\lambda_1\frac{\dot a}{a}R \nonumber \\
&&-\gamma'\lambda_1\left(12\frac{\dot a}{a}\dot R +R(4R^0_0-R)\right).
\end{eqnarray}
Again, in order to get the corresponding left-hand-sides, one should put $\lambda_1\equiv1$ and erased ``primes''.

At last, let us write down the complete set of the field equations.

Vector:
\begin{equation}\label{6gamma}
-6\gamma\dot R=G^0.
\end{equation}
Tensor:
\begin{equation}\label{tensor1}
-\gamma\left(12\frac{\dot a}{a}\dot R+ R(4R^0_0-R)\!\right)=T^0_0,
\end{equation}
\begin{equation}\label{tensor2}
-12\gamma\left(\ddot R+ 3\frac{\dot a}{a}\dot R\!\right)=T.
\end{equation}
Self-consistency condition:
\begin{equation}\label{Self-c}
2\frac{(G^0a^3)^{\dot{}}}{a^3}=T^0_0+3T^1_1=T,
\end{equation}
where
\begin{equation}\label{R002}
R^0_0=-3\frac{\ddot a}{a},
\end{equation}
\begin{equation}\label{R7}
R=-6\left(\frac{\ddot a}{a}+\frac{\dot a^2+k}{a^2}\right), \quad k=0,\pm1
\end{equation}
It is quite clear that the self-consistency  condition is just the consequence of the vector  and trace equations.

How about the equations of motion for the cosmological perfect fluid?	
It appears that we are left with only one equation plus the law of the particle creation, namely,
\begin{equation}\label{dotlambda}
\left\{
\begin{array}{l}	
\dot\lambda_1=-\frac{\varepsilon+p}{n} \\
\frac{(na^3)^{\dot{}}}{a^3}=\Phi({\rm inv})
\end{array}	
\right.
\end{equation}
\begin{equation}\label{Phiinv}
\Phi({\rm inv})=-\frac{4}{3}\beta'R^0_0(2R^0_0-R)+\gamma'R^2.
\end{equation}
Remember, we assumed the absence of any classical fields that may cause the particle production.

Let us suppose now that the universe was created from ``nothing'', i.\,e., from some quantum foam\cite{Vilenkin}. Most likely, it emerged empty --- without matter fields (particles). This vacuum is not absolutely empty, it is filled with the quantum (virtual)  fluctuations of the matter fields (particles are the real quanta of these fields). 

Such a vacuum, being deformed (polarized) by the strong gravitational field, can produce particles. Then, the question arises: being the initial state, may or may not it persists? In other words, may it survive for a while? If the answer is ``yes'', we will call this physical vacuum ``the pregnant'' vacuum --- it may give birth to particles, but did not do this yet (see also\cite{bde19,bdes20}).

What are the condition to be the pregnant vacuum? First of all, 
\begin{equation}\label{Phiinv0}
\Phi({\rm inv})=0, \quad |\beta'|+|\gamma'|\neq0,
\end{equation}
i.\,e., the particle creation is not completely prohibited. This is translated to
\begin{equation}\label{43}
\frac{4}{3}\beta'R^0_0(2R^0_0-R)=\gamma'R^2.
\end{equation}
Then, evidently,
\begin{equation}\label{n0}
n=0
\end{equation}
and
\begin{equation}\label{G0part0}
G^0{[\rm part]}=T^0_0{[\rm part]}=T{[\rm part]}=0.
\end{equation}
Let us have a look at the equation left from the set of equations of motion for the perfect fluid, namely
\begin{equation}\label{dotlambda2}
\dot\lambda_1=-\frac{\varepsilon+p}{n}.
\end{equation}
There are three completely different types of behavior of the right-hand-side in the limit $n\to0$.
\begin{enumerate}
\item $\lim\limits_{n\to0}\frac{\varepsilon+p}{n}=\infty$. There is no vacuum solution at all, even in the very  beginning.                                                                                                                                                                                                             
\item $n\to0$ 
$\lim\limits_{n\to0}\frac{\varepsilon+p}{n}=0$ (for example, thermal radiation). Then, if $\beta',\gamma'\neq0$, we have the following set of equations,
\begin{equation}\label{lambda1c}
\lambda_1=const
\end{equation}
\begin{equation}\label{Rxi}
R=\xi R^0_0 \quad  \Rightarrow
\end{equation}
\begin{equation}\label{Rxi2}
(3\gamma'\xi^2+4\beta'(\xi-2))R^0_0=0
\end{equation}
\begin{equation}\label{6gamma2}
-6(\gamma-\gamma'\lambda_1)\dot R=0
\end{equation}
\begin{equation}\label{12gamma}
-12(\gamma-\gamma'\lambda_1)\frac{(\dot Ra^3)^{\dot{}}}{a^3}=0
\end{equation}
\begin{equation}\label{6gamma2b}
-(\gamma- \gamma'\lambda_1)\left\{12\frac{\dot a}{a}\dot R 
+R(R-4R^0_0)\right\}=0.
\end{equation}
Let, first, $\gamma\neq\gamma'\lambda_1$, then $\dot R=0$. Now we have two possibilities. Either $R=0$ and, hence, $R^0_0=0$ --- in this case the vacuum represents the so called  Milne universe, i.\,e., locally flat (Minkowski) space-time. Or $\xi=4$, one gets the (Anti) de Sitter  space-time for the specific choice of the coefficients $\beta+6\gamma'=0$.

Let now	
\begin{equation}\label{gamma3}
\gamma=\gamma'\lambda_1,
\end{equation}
it is not the special condition, but the solution for $\lambda_1$. Then
\begin{equation}\label{dota2}
\dot a^2+k=C_0 a^{\frac{4}{\xi-2}}.
\end{equation}
Formally the vacuum solution exists, but, first, it may be unstable, and, second, if no other types of particle are producing.

At last, let us come to the third case.

\item Dust pregnancy. For the dust matter
\begin{equation}\label{dota2b}
\lambda_1=-\phi(0)(t-t_0).
\end{equation}
The whole set of equations are reduced to
\begin{equation}\label{dotR}
\dot R=0
\end{equation}
\begin{equation}\label{dotR2}
R(R-4R^0_0)=0.
\end{equation}
Up to now, the results are the same as before. But, for the general choice of $\beta'$ and $\gamma'$, the universe emerging from the quantum foam, like Aphrodite, immediately starts to produce dust particles!!! Dark matter? 
\end{enumerate}

\end{document}